\documentclass{aa}
\usepackage{graphicx}
\usepackage{txfonts}
\usepackage{natbib}
\bibpunct{(}{)}{;}{a}{}{,}    

\def\me{$M_{\oplus}$}
\def\rmh{$R_{\rm{mH}}$}

\begin{document}

\title
{Three-dimensional simulations of multiple protoplanets embedded in a 
protostellar disc}

\author{P.\ Cresswell and R.~P.\ Nelson}

\institute{ Astronomy Unit, 
 Queen Mary, University of London, Mile End Rd, London, E1 4NS, U.K.}
 
\offprints{P.Cresswell@qmul.ac.uk}

\date{Received December 2007 /Accepted February 2008}

\def\LaTeX{L\kern-.36em\raise.3ex\hbox{a}\kern-.15em
         T\kern-.1667em\lower.7ex\hbox{E}\kern-.125emX}



\authorrunning{P.\ Cresswell \& R.~P.\ Nelson}
\titlerunning{Multiple protoplanets in a protostellar disc.}

\abstract
{Theory predicts that low-mass protoplanets in a protostellar 
disc migrate into the central star on a time scale that is short
compared with the disc lifetime or the giant planet formation time scale.
Protoplanet eccentricities of $e\ga H/r$ can slow or reverse migration,
but previous 2D studies of multiple
protoplanets embedded in a protoplanetary disc have shown that
gravitational scattering cannot maintain significant planet eccentricities
against disc-induced damping. The eventual fate of these systems
was migration into the central star.}
{Here we simulate the evolution of low-mass
protoplanetary swarms 
in three dimensions. The aim is to examine both protoplanet survival rates 
and the dynamical structure of the resulting planetary systems, and to compare
them with 2D simulations.}
{We present results from a 3D hydrodynamic simulation of
eight protoplanets embedded in a protoplanetary disc. We also
present a suite of simulations performed using an $N$-body
code, modified to include prescriptions for planetary migration and for
eccentricity and inclination damping. These prescriptions were obtained
by fitting analytic formulae to hydrodynamic simulations
of planets embedded in discs with initially eccentric and/or inclined
orbits.}
{As was found in two dimensions, differential migration produces groups of 
protoplanets in stable, multiple mean-motion resonances that migrate in 
lockstep, preventing prolonged periods of gravitational scattering. 
In almost all simulations, this leads to large-scale migration of
the protoplanet swarm into the central star in the absence of a viable
stopping mechanism. The evolution involves mutual collisions,
occasional instances of large-scale scattering, and the frequent formation of
the long-lived, co-orbital planet systems that arise in $> 30$\% of all runs.}
{Disc-induced damping overwhelms eccentricity and inclination growth due to 
planet-planet interactions, leading to large-scale migration of
protoplanet swarms. Co-orbital planets are a natural outcome of
dynamical relaxation in a strongly dissipative environment, and if observed
in nature would imply that such a period of evolution commonly arises
during planetary formation.}


\maketitle

\section{Introduction}
\label{intro}

The observed lower limits on extrasolar planet mass are continuing
to decrease, while at the same time the number of known multiple 
planet systems is continuing
to grow \citep{rivera05, lovis06, udry07}. Multiple planet systems
containing sub-Neptune mass planets are now being discovered.
Missions such as CoRoT and Kepler are expected to
discover further sub-Neptune mass planets beyond the reach of current 
observations, leading the way toward finding planetary systems more 
like our own.

Formation of a planetary system is believed to involve 
accretion within a protoplanetary disc, involving essentially three
steps: coagulation of dust grains into small ($\sim$ 1 km) planetesimals; 
runaway growth of planetesimals into larger ($\sim 100$--$1000$ km) 
protoplanets; oligarchic growth by planetesimal accretion into larger 
planetary cores.
Those cores forming beyond the snow line are expected to reach masses of
$\sim 10$ \me, accreting a gaseous envelope to become gas giants if they
form before disc dispersal, or ice giants should they form late in the
disc's lifetime \citep{bodenheimer86, pollack96}. Smaller, Mars-mass bodies 
result from oligarchic growth in the terrestrial zone, and
accrete {\it via} giant impacts to form an inner system of rocky, 
terrestrial-mass planets \citep{chambers98}.

One of several problems associated
with this picture is the rapid inward migration
experienced by a protoplanet due to the gravitational interaction between it
and the gaseous disc \citep{ward97, tanaka02}.
In particular, understanding the formation of 
giant planets, which must spend more that $10^6$ years
in the disc in order to 
accrete a gas envelope, remains an unsolved problem.
Their solid cores have
migration times shorter than both the gas accretion time scale
and the disc life time \citep{papaloizou05}.
Referred to as type I migration, this drift makes it hard to understand how
gas giants can form at all without being accreted by the central star. 
Solving the type I problem remains an active area of research,
and recently suggested remedies include: stochastic migration in
a turbulent disc \citep{nelson04, nelson05};
corotation torques operating in a region of positive gradient in
disc surface density \citep{masset06};
corotation torques in radiatively inefficient discs 
\citep{paardekooper06}. In this paper we further explore the role
of protoplanet scattering (\cite{cresswell06} --- hereafter Paper I).

Many authors have examined the interactions of multiple planetary embryos, or 
fully formed planets, within protoplanetary discs. Numerical simulations of the
oligarchic growth phase have been used to examine the interactions of
planetary embryos \citep{kokubo00, thommes03}.
\citet{mcneil05} found that a proto-terrestrial system may form against
type I migration by enhancing the disc mass by a factor of 2--4.
Resonant capture between 
two planets in the 1--20 \me~range has been studied by \citet{papaloizou05b}.
\citet{thommes05} suggests that the formation of the first giant planet may
be a significant step in the formation of a planetary system, by capturing
smaller cores in resonance and preventing further type I migration. Previously,
pairs of giant planets in resonance have been examined,
often with direct application to a specific system such as GJ 876
\citep{snellgrove01, kley05}.

One area that has not been addressed to the same depth is the issue of how a
swarm of protoplanetary cores, of Earth mass and above, will evolve under the
influence of a surrounding protoplanetary disc. Models of oligarchic growth
\citep{kokubo00, thommes03} predict that a number of cores should form coevally,
separated in radius by $\sim8$ mutual Hill radii. 
Differential type I migration may 
cause these bodies to undergo close encounters, leading to gravitational 
scattering and the pumping of eccentricities.

\citet{papaloizou00} found that 
type I migration can be slowed or reversed when a protoplanet embedded in a 
disc achieves an eccentricity $e \ga 1.1H/r$, where $H/r$ is the disc's 
scale height-to-radius ratio, raising the possibility of the mutual 
interactions of a swarm of cores sustaining significant eccentricities and 
slowing type I migration for at least some of them.
In an earlier study, however, 
using 2D $N$-body and hydrodynamic models we found that the damping action of
the disc is too strong to sustain eccentricities in this manner (see Paper I).
Instead 
the cores form resonant groups and migrate inwards in lockstep. Due to
the 2D modelling it is possible that the collision probability was 
overestimated, 
removing a higher number of protoplanets than should be properly expected,
and the influence of planetary inclinations on reducing the migration rate
was neglected.

To address these issues, we perform 3D numerical simulations of swarms of 
planetary cores embedded in a protoplanetary disc. In the first instance
we  present the results of a full, 3D hydrodynamic simulation
of a protoplanetary disc containing eight protoplanets.
This simulation provides a good illustration of the early stages
of evolution, but cannot be run for long times. To examine the long term
evolution, we also present 3D simulations performed with an $N$-body code,
modified to include prescriptions for migration, and eccentricity
and inclination damping. These prescriptions were obtained by
fitting analytic formulae to numerous 3D hydrodynamic simulations
of protoplanets on eccentric and/or inclined orbits embedded in a
protoplanetary disc.
The results of these 3D multiplanet simulations suggest that gravitational
interactions among a swarm remains ineffective at maintaining the requisite 
eccentricities to slow/stop migration, due to the strong damping from the disc.
We also find that many co-orbital planets form naturally from such a migrating
population, raising the possibility of their detection among the observed
extrasolar hot Neptunes and super-earths.

The plan of this paper is as follows. In Sect.~2 we describe the basic
equations of 
motion. In Sect.~3 we describe the two numerical schemes and provide equations
describing the disc's action on the protoplanets. In Sect.~4 we describe the
initial conditions. In Sect.~5 we present a hydrodynamic
multiple-planet model. In Sect.~6 we present the results obtained from the 
modified $N$-body scheme, and discuss the trends and implications of these 
results. We give our conclusions in Sect.~7.

\section{Equations of Motion} \label{eqns} 

An unperturbed protoplanetary disc 
with constant aspect ratio $H/r$ can be conveniently described using
spherical polar coordinates ($r,\theta, \phi$) with the
origin located at the central star. The continuity equation is expressed as
\begin{equation}
\label{cont-eq}
\frac{\partial \rho}{\partial t} 
  + \nabla \cdot \left ( \rho \mbox{\boldmath$v$} \right ) = 0,
\end{equation}
and the three components of the momentum equation are written
\begin{equation}
\label{mom-r-eq}
\frac{\partial \left ( \rho v_r \right )}{\partial t} 
    + \nabla \cdot \left ( \rho v_r \mbox{\boldmath$v$} \right )
  =   \frac{\rho (v_{\theta}^2 + v_{\phi}^2)}{r}   
    - \frac{\partial p}{\partial r} 
    - \rho \frac{\partial \Phi}{\partial r}
    + f_r 
\end{equation}
\begin{eqnarray}
\label{mom-theta-eq}
\frac{\partial \left ( \rho v_{\theta} \right )}{\partial t}
   + \nabla \cdot \left ( \rho v_{\theta} \mbox{\boldmath$v$} \right ) 
  & = & -\frac{\rho v_r v_{\theta}}{r} + \frac{\rho v_{\phi}^2 \cot \theta}{r} \nonumber \\
  &   & - \frac{1}{r} \frac{\partial p}{\partial r}
        - \frac{\rho}{r} \frac{\partial \Phi}{\partial \theta}
        + f_{\theta}   
\end{eqnarray}
\begin{eqnarray}
\label{mom-phi-eq}
\frac{\partial \left ( \rho v_{\phi} \right )}{\partial t}  
     + \nabla \cdot \left ( \rho v_{\phi} \mbox{\boldmath$v$} \right )
  & = & -\frac{\rho v_r v_{\phi}}{r} 
              - \frac{\rho v_{\theta} v_{\phi} \cot \theta}{r} \nonumber \\
  &   & - \frac{1}{r \sin \theta} \frac{\partial p}{\partial \phi}
        - \frac{\rho}{r \sin \theta} \frac{\partial \Phi}{\partial \phi}
        + f_{\phi}    
\end{eqnarray}
Here $\rho$ denotes the density, $p$ is the gas pressure and $f_r$,
$f_{\theta}$, $f_{\phi}$ are the viscous forces per unit volume in the radial,
meridional and azimuthal directions, respectively. $v_r$, $v_{\theta}$
and $v_{\phi}$ denote the corresponding velocities. The gravitational potential
$\Phi$ is given by
\begin{eqnarray}
\Phi(\mbox{\boldmath$r$}) 
 & = & - \frac{G M_*}{r}
         - \sum_{p=1}^N{\frac{G m_p}{\sqrt{ | \mbox{\boldmath$r$} - \mbox{\boldmath$r$}_p |^2 + \epsilon^2} } }  \nonumber \\
 &   &   + \sum_{p=1}^N{\frac{G m_p}{r_p^3}\mbox{\boldmath$r$}\cdot \mbox{\boldmath$r$}_p}
    + G \int_{\rm{V}}{\frac{\rm{d} m(\mbox{\boldmath$r$})}{r^{3}}\mbox{\boldmath$r$}\cdot \mbox{\boldmath$r$}_p},  
\label{grav-pot-eq} 
\end{eqnarray}
where $M_*$ is the stellar mass, $\epsilon$ is a softening parameter, and the
summations are over all protoplanets $p$ with masses $m_p$. The subscript `$p$'
denotes evaluation at the location of the protoplanet. The latter two terms in
Eq.~\ref{grav-pot-eq} result from acceleration of the coordinate system
due to the gravity of the protoplanets and the protostellar disc. The integral
is performed over the volume of the disc.

Each protoplanet experiences the gravitational acceleration from the central 
star, the other protoplanets and the protostellar disc. The equation of
motion for each protoplanet is:
\begin{eqnarray}
\frac{\rm{d}^2 \mbox{\boldmath$r$}_p}{\rm{d} t^2}
 & = & - \frac{G M_*}{r^3_p} \mbox{\boldmath$r$}_p
            - \sum_{p' \not= p}^N{ \frac{G m_{p'}} {| \mbox{\boldmath$r$}_p 
             - \mbox{\boldmath$r$}_{p'} |^3} (\mbox{\boldmath$r$}_p - \mbox{\boldmath$r$}_{p'}) } \nonumber \\
 &   &  - \nabla \Phi_{\rm{d}}
          - \sum_{p'=1}^{N}{\frac{G m_{p'}}{r_{p'}^3} \mbox{\boldmath$r$}_{p'} }
\label{planet-eom} 
\end{eqnarray}
where
\begin{equation}
\label{disc-pot-eq}
\Phi_{\rm{d}}(\mbox{\boldmath$r$}_p)
  = - G \int_{\rm{V}} 
        \frac{\rho(\mbox{\boldmath$r$}) \rm{d}\mbox{\boldmath$r$}}
                 {\sqrt{ | \mbox{\boldmath$r$} - \mbox{\boldmath$r$}_p |^2 + \epsilon^2} }
    + G \int_{\rm{V}}
        \frac{\rm{d} m(\mbox{\boldmath$r$})}
                 {r^{3}}
       \mbox{\boldmath$r$}_p\cdot \mbox{\boldmath$r$}
\end{equation}
is the gravitational potential due to the disc. The integrals are again performed
over the volume of the disc. The final terms in Eqs.~\ref{planet-eom} and
\ref{disc-pot-eq} are the indirect terms arising from the acceleration of the 
coordinate system by the protoplanets and disc, respectively.

We adopt 
a locally isothermal equation of state such that the
pressure and density are related by
\begin{equation}
\label{eq-of-state}
p=c_{\rm{s}}^2 \rho
\end{equation}
where $c_{\rm{s}} = \left ( \frac{H}{r} \right ) v_{\rm{K}}$ is the
isothermal sound speed, a fixed function of distance from the central star.
$v_{\rm{K}}$ is the local Keplerian velocity. All disc models have aspect ratio
$H/r=0.05$. We also adopt the alpha model for the disc viscosity such
that the kinematic viscosity $\nu=\alpha c_s H$, with 
$\alpha=5 \times 10^{-3}$.

\section{Numerical methods}
\label{methods}          

We use two distinct numerical schemes: a 3D hydrodynamic disc model 
together with embedded planets is computed using a hydrodynamics code
(NIRVANA); a suite of simulations are computed
using a much faster $N$-body code which has been adapted to
emulate the effects of orbital migration, and eccentricity and inclination 
damping on the protoplanets due to the protoplanetary disc. We describe each in
turn in the following sections, and demonstrate that the modified
$N$-body code agrees well with results from the hydrodynamic code.

\subsection{Hydrodynamic scheme}
\label{hydro-method}

We use a modified version of the grid-based code NIRVANA to conduct 3D
hydrodynamical simulations \citep{zeigler97}. The code has previously been 
applied to a variety of disc-planet numerical studies in both two and three
dimensions. Further details may be found in \citet{nelson00}, \citet{cresswell06b},
\citet{cresswell07} and in Paper I.

The motion of the protoplanets is integrated using a $5^{\rm{th}}$-order
Runge-Kutta scheme \citep{press92}. As in Paper I disc self-gravity is 
neglected.
We employ reflecting boundary conditions in the meridional direction,
and wave-damping conditions at the inner and outer radial boundaries.
The implementation of these is described in Paper I, along with a
description of the time step control procedure.

In common with many hydrodynamical disc simulations 
and \citet{cresswell07},
we adopt a scale height $H/r=0.05$ and surface density profile $\Sigma \propto
r^{-0.5}$. A disc opening half-angle of $10^{\circ}$ then models 
three and a half scale heights.
Our adoption of a locally isothermal equation of state means that 
wave propagation is primarily confined to the radial direction, such
that the use of reflecting boundary conditions at the meridional boundaries
does not lead to significant wave reflection toward the disc midplane.
The development of inclined orbits for the protoplanets, however,
can cause the excitation of bending waves in the disc, which could in
principle be affected by the vertical boundary conditions.
Test simulations of inclined planets in discs using zero-gradient outflow
boundary conditions, hwoever, indicate that the results presented
in this paper are not strongly affected by the choice of meridional
boundary conditions. The adoption of reflecting conditions prevents
the slow loss of mass from the disc which accompanies the use
of open boundaries.

\subsubsection{Code resolution and modelling}
\label{disc-dynamics}

Due to the large computational expense of
3D disc-planet studies, the spatial resolution used in each model must
be carefully weighed against the duration and number of simulations required.
In \citet{cresswell07} it was found that the coarsest resolution tested 
produced results of a similar character in migration rate
and eccentricity and inclination damping rates to the highest resolutions 
tested, with some difference in absolute migration/damping times; numerical
convergence of the code was established with these high-resolution models.
Since we have needed to run many simulations to obtain fitting
formulae for the modified $N$-body code, we have chosen a resolution for
those simulations
that leads to accurate results, but which makes the problem tractable:
$(N_r,N_{\theta},N_{\phi}) = (128,40,300)$. This resolution corresponds
closely to the lower-resolution runs presented in \citet{cresswell07}.
The limits of the computational domain are defined by the intervals
$r=[0.4,2.5], \theta=[80^{\circ},100^{\circ}], \phi=[0,2\pi]$. 
With the set-up described, we follow the evolution of a 10 Earth 
mass (\me) planet on a variety of eccentric and inclined orbits,
and fit the migration and eccentricity and inclination damping rates
using simple nonlinear functions of $e$ and $i$.

We also performed one hydrodynamic
simulation consisting of eight protoplanets embedded
in a protoplanetary disc. The resolution adopted for this simulation
was $(N_r,N_{\theta},N_{\phi}) = (288,40,444)$, with the limits of
the computational domain given by 
$r=[0.6,3.0], \theta=[80^{\circ},100^{\circ}], \phi=[0,2\pi]$.

\subsection{$N$-body scheme}
\label{nbody-method}

The second method we use to evolve a disc-planet system is an $N$-body 
integrator (to which we apply the name HENC-3D) with analytic
functions which model the effects of type I 
migration, and eccentricity and inclination damping. The integrator used is the
same $5^{\rm{th}}$-order Runge-Kutta routine as used in NIRVANA. To prevent the
time step from becoming too low, protoplanets were removed from the simulation 
if they fell within 10 Solar radii of the origin.

Due to the computational expense of 3D hydrodynamic computations, this is our
primary method of evolving multiple-planet systems; even if it were practical
to model a disc in NIRVANA over the scales required, HENC-3D is $\sim 10^7$ 
times faster at evolving such systems even at the moderate resolution
we employ.

\subsubsection{Prescriptions for eccentricity and inclination damping
and migration}
\label{damping}

We construct prescriptions for the migration rates
and eccentricity and inclination
damping rates which are incorporated into the $N$-body code. These allow us to
follow the evolution of protoplanet swarms for longer
than is possible with the hydrodynamic code.
Our starting point is the formulae for damping rates
obtained by \citet{tanaka04} and the migration rates obtained
by \citet{tanaka02}, which we modify using multiplicative factors
that account for changes in damping and migration rates due to
planetary eccentricity and/or inclination. These multiplicative
factors are obtained by fitting formulae to the results of
numerous hydrodynamical simulations, performed with different
values of $e$ and $i$. The results of a subset of these simulations
are shown in Figs.~1--3. In the case of eccentricity and
inclination damping, the formulae are founded on the
damping time scale derived by \citet{tanaka04}, and given by their
Eq.~49:
\begin{equation}
\label{twave-eq}
t_{\rm{wave}} = \frac{M_*}{m_p}
                \frac{M_*}{\Sigma_p a_p^2}
                \left ( \frac{H}{r} \right )^4 \Omega_p^{-1}
\end{equation}
where $\Omega = v_{\phi}/r$ is the angular velocity of the unperturbed disc and
$\Sigma$ is the surface density
\begin{equation}
\label{surfdens-eq}
\Sigma=\int_{-\infty}^{\infty} \rho~\rm{d}z.
\end{equation}
The eccentricity damping time that best fits our simulation
results is given by:
\begin{eqnarray}
\label{ecc-eq}
t_{\rm{e}} & = & \frac{t_{\rm{wave}}}{0.780} \times \\
           &   & \left [ 1 - 0.14 \left ( \frac{e}{H/r} \right )^2
                        + 0.06 \left ( \frac{e}{H/r} \right )^3
                        + 0.18 \left ( \frac{e}{H/r} \right )
                               \left ( \frac{i}{H/r} \right )^2 \right ].
                                           \nonumber
\end{eqnarray}
where the terms in square brackets provides the modification to
the damping rate caused by the finite eccentricity and/or inclination.
As was found by \citet{papaloizou00} and \citet{cresswell07}, our results
show that
$de/dt \propto e^{-2}$ at high $e$, with exponential decay for $e< H/r$.
The inclination damping time, $t_{\rm i}$, is given by:
\begin{eqnarray}
\label{inc-eq}
t_{\rm{i}} & = & \frac{t_{\rm{wave}}}{0.544} \times \\
           &   &  \left [ 1 - 0.30 \left ( \frac{i}{H/r} \right )^2
                            + 0.24 \left ( \frac{i}{H/r} \right )^3
                            + 0.14 \left ( \frac{e}{H/r} \right )^2
                                   \left ( \frac{i}{H/r} \right )   \right ].
                                          \nonumber
\end{eqnarray}
where again the terms in square brackets modify the damping rate
because of finite eccentricity and/or inclination.
As described in \citet{cresswell07} we find
$di/dt \propto i^{-2}$ at high $i$, with exponential decay for $i< H/r$.

The prescription for the migration time, $t_{\rm m}$,
is based on the results of 
\citet{tanaka02}, who considered circular orbits only. 
The results of our simulations are well-fitted by the the
expression:
\begin{eqnarray}
\label{mig-eq}
t_{\rm{m}} & = & \frac{2 t_{\rm{wave}}}{2.7+1.1\beta} 
                 \left ( \frac{H}{r} \right )^{-2}  
                 \left [ P(e) + \frac{P(e)}{|P(e)|} \right. \times \\
&  & \left.  \left \{ 0.070 \left ( \frac{i}{H/r} \right )
                          + 0.085 \left ( \frac{i}{H/r} \right )^4
                          - 0.080 \left ( \frac{e}{H/r} \right )
                                  \left ( \frac{i}{H/r} \right )^2 
                          \right \}
                            \right ]   \nonumber  
\end{eqnarray}
where
\begin{displaymath}
P(e) = \frac{1 + \left ( \frac{e}{2.25H/r} \right )^{1.2}
                   + \left ( \frac{e}{2.84H/r} \right )^6}
                {1 - \left ( \frac{e}{2.02H/r} \right )^4 }
\end{displaymath}
and $\beta$ is given by $\Sigma(r) \propto r^{-\beta}$.
The sign dependency on $P(e)$ allows 
torque reversal at sufficiently high $e$.
\begin{figure}[t]
\begin{center}
\includegraphics[width=7.5cm]{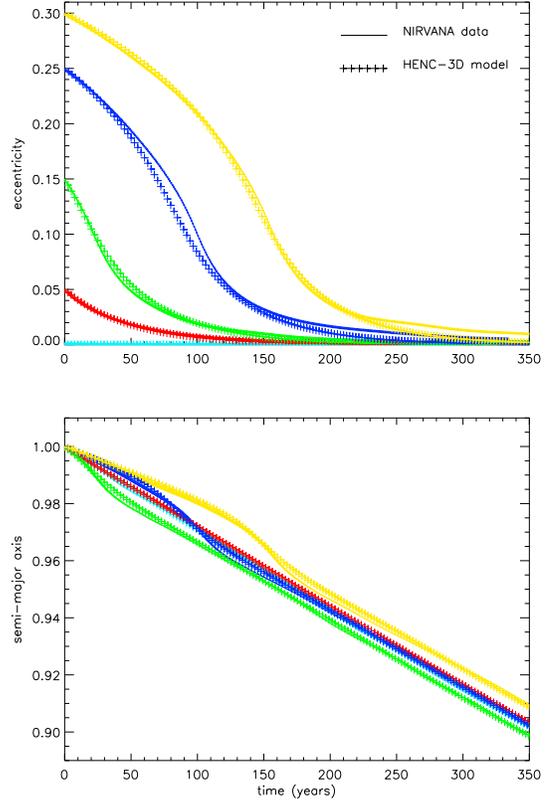}
\caption{({\it Top}) Eccentricity evolution of a $10$ \me~
protoplanet on a variety of eccentric orbits. Solid lines are the results
from NIRVANA, crosses from the $N$-body code. 
({\it Bottom}) The corresponding migration rates.}
\label{fig1}
\end{center}
\end{figure}
We note that \citet{papaloizou00} undertook a study of the 
disc-induced migration and eccentricity damping of low-mass 
protoplanets using linear theory, and derived similar expressions
to those given in Eqs.~\ref{ecc-eq} and \ref{mig-eq}. Our expressions
differ from theirs only because we found that Eqs.~\ref{ecc-eq}
and \ref{mig-eq} better fit our hydrodynamical simulations,
and we consider both eccentric and inclined orbits.
\begin{figure}[t]
\begin{center}
\includegraphics[width=7.5cm]{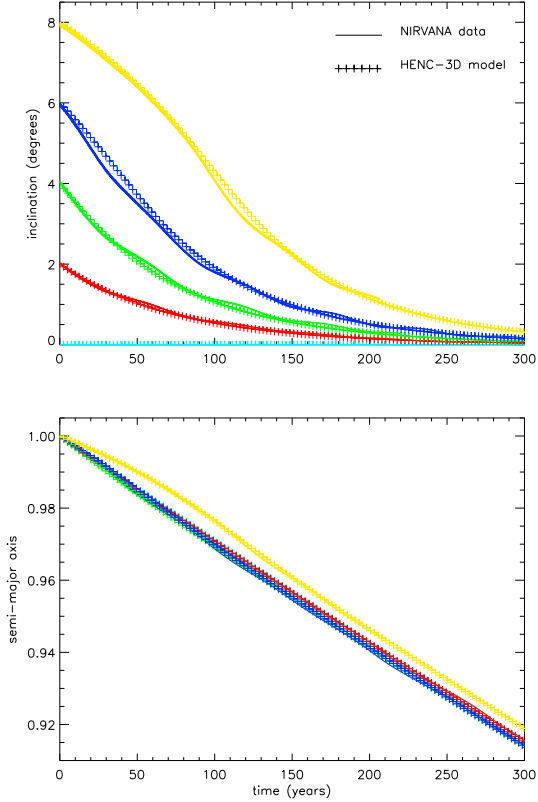} 
\caption{({\it Top}) Inclination evolution of a $10$ \me~
protoplanet on a variety of inclined orbits. Solid lines are the results
from NIRVANA, crosses from the $N$-body code. 
({\it Bottom}) The corresponding migration rates.}
\label{fig2}
\end{center}
\end{figure}

We implement the following expressions in the $N$-body code as accelerations
experienced by the protoplanets due to the disc, using values of $t_{\rm{m}}$,
$t_{\rm{e}}$ and $t_{\rm{i}}$ obtained from Eqs.~\ref{ecc-eq}--\ref{mig-eq}:
\begin{equation}
\label{macc-eq}
\mbox{\boldmath$a$}_{\rm{m}} = - \frac{\mbox{\boldmath$v$}}{t_{\rm{m}}},
\end{equation}
\begin{equation}
\label{eacc-eq}
\mbox{\boldmath$a$}_{\rm{e}} = 
       -2 \frac{(\mbox{\boldmath$v$}.\mbox{\boldmath$r$})\mbox{\boldmath$r$} }
               {r^2 t_{\rm{e}}},
\end{equation}
\begin{equation}
\label{iacc-eq}
\mbox{\boldmath$a$}_{\rm{i}} = - \frac{v_z}{t_{\rm{i}}}
                                 \mbox{\boldmath$k$},
\end{equation}
where $\mbox{\boldmath$k$}$ is the unit vector in the $z$-direction.

\subsubsection{Comparison between $N$-body and hydrodynamic code }
\label{agreement}

We briefly demonstrate the agreement between the modified $N$-body 
and hydrodynamic codes. Figure \ref{fig1} shows the orbital evolution 
of a planet on a variety of eccentric orbits, and the corresponding 
planetary migration. The decay of eccentricity at both high and low 
eccentricity is modelled accurately,
with the largest deviation at the transition from 
quadratic to exponential decay.
We find that the exchange of angular momentum between disc and
planet reverses sign when $e \ga 2H/r$, rather than the value
$e \ga 1.1H/r$ reported by \citet{papaloizou00}.
In agreement with \citet{cresswell07},
we find the peak {\it positive} angular
momentum exchange rate (from the disc to the planet) to occur when 
$e\sim 4 H/r$, rather 
than the value $e \sim 2H/r$ reported by \citet{papaloizou00}.
This may be an effect of adopting a different surface density profile.
\begin{figure}[t]
\begin{center}
\includegraphics[width=7.5cm]{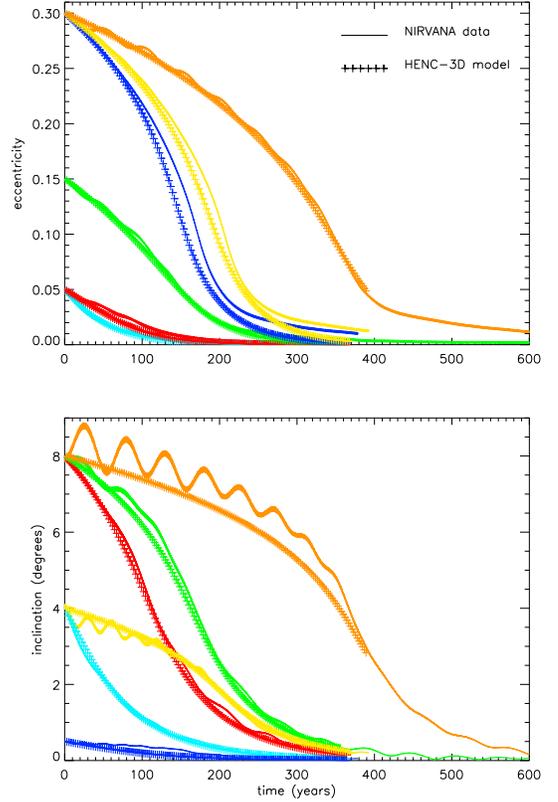}
\caption{({\it Top}) Eccentricity evolution of a $10$ \me~
protoplanet on a variety of eccentric and inclined orbits. Solid lines are the
results from NIRVANA, crosses from the $N$-body code.
({\it Bottom}) Inclination evolution of the same orbits. }
\label{fig3}
\end{center}
\end{figure}

Figure \ref{fig2} shows the orbital evolution of a planet on a variety of
inclined orbits, and the accompanying migration. As with the eccentric orbits,
agreement between the hydrodynamic model and the modified $N$-body code
is generally good, and poorest at the transition to exponential decay.

Figure \ref{fig3} shows the orbital evolution of a planet on a variety of
orbits which are both eccentric and inclined to the disc midplane. Agreement
between the two numerical schemes is again good overall.
The oscillations present in the most strongly excited runs are physical
artefacts, and possess a frequency related to the precession rate of the 
longitude of ascending node. A bending distortion of the disc by the protoplanet is 
probably responsible, but does not significantly influence the underlying damping rate.
It should be noted that our prescription for inclination evolution given by 
Eq.~\ref{inc-eq} does not include this oscillatory behaviour.

We note that it is not feasible to test planetary orbits with very
high eccentricity or inclination against hydrodynamic models, due to the 
large range of scales over which the disc model must be simulated. However, 
previous experiments \citep{cresswell06b} and Paper I imply that swarms of 
protoplanets will 
spend the majority of their time with fairly low eccentricities and
inclinations.
Equations \ref{ecc-eq}--\ref{mig-eq} successfully capture the character of
the dynamical evolution 
($dy/dt \propto t$ at low $y$ and $\propto y^{-2}$ at $y \gg H/r$, for $y=e$ or $i$)
during the brief times when these quantities become large.

\section{Initial conditions and units}
\label{set-up}

We adopt a similar strategy to Paper I when setting up planetary initial
conditions. We first define the semi-major axis of
the innermost body to be $a_1 = 5$ AU, such that the population exists beyond
the snow line where the proportion of solids available for protoplanet core
formation is higher. Successive protoplanet
semi-major axes were determined by choosing their separations to be a
specified number of mutual Hill radii. Thus 
$a_{i+1} = a_i + N_{\rm{mH}}R_{\rm{mH}}$ where $N_{\rm{mH}}$ is typically 5
(though other values are considered). The mutual Hill radius is defined by
\begin{equation}
\label{rmh-eq}
R_{\rm{mH}} = \left ( \frac{m_i+m_j}{3M_*} \right )^{\frac{1}{3}}
              \left ( \frac{a_i + a_j}{2} \right ).
\end{equation}
For two planets on initially circular orbits, rapid instability occurs if the
separation $\Delta$ between planets is less than the critical value
\begin{equation}
\label{deltacrit-eq}
\frac{\Delta_{\rm{crit}}}{R_{\rm{mH}}} = 2\sqrt{3} \approx 3.46
\end{equation}
\citep{gladman93, chambers96}. Simulations of oligarchic growth 
\citep{kokubo00, thommes03} suggest that the mutual separation of protoplanets
is normally $\simeq 8~R_{\rm{mH}}$. We typically adopt smaller spacings to
maximise close encounters, although larger separations are also investigated.

\begin{table}
\begin{center}
\begin{tabular}{|c|c|c|c|c|c|c|c|c|}
\hline
Set & $\mu_e$ & $\sigma_i$ & $\mu_m$ & $\sigma_m$ & Cutoff & $\Delta_0$
        & $\Sigma$ & $N_p$ \\ \hline
H1 & 0.05 & $2^{\circ}$ & - & - & - & 5 & 1 & 8 \\ \hline
O1 & 0.05 & $2^{\circ}$ & - & - & - & 5 & 1 & 10 \\ \hline
O2 & 0.15 & $2^{\circ}$ & - & - & - & 5 & 1 & 10 \\ \hline
O3 & 0.05 & $6^{\circ}$ & - & - & - & 5 & 1 & 10 \\ \hline
O4 & 0.10 & $4^{\circ}$ & - & - & - & 5 & 1 & 10 \\ \hline
O5 & 0.05 & $2^{\circ}$ & - & - & - & 4 & 1 & 10 \\ \hline
O6 & 0.05 & $2^{\circ}$ & - & - & - & 8 & 1 & 10 \\ \hline
O7 & 0.05 & $2^{\circ}$ & - & - & - & 5 & 0.5 & 10 \\ \hline
O8 & 0.05 & $2^{\circ}$ & - & - & - & 5 & 0.2 & 10 \\ \hline
O9 & 0.05 & $2^{\circ}$ & - & - & - & 5 & 1 & 5 \\ \hline
R1 & 0.05 & $2^{\circ}$ & 10 & 7 & 0.2 & 5 & 1 & 10 \\ \hline
R2 & 0.05 & $2^{\circ}$ & 10 & 5 & 2 & 5 & 1 & 10 \\ \hline
R3 & 0.05 & $2^{\circ}$ & 10 & 3 & 2 & 5 & 1 & 10 \\ \hline
R4 & 0.05 & $2^{\circ}$ & 7 & 3 & 2 & 5 & 1 & 10 \\ \hline
R5 & 0.05 & $2^{\circ}$ & 7 & 5 & 0.2 & 5 & 1 & 10 \\ \hline
R6 & 0.10 & $2^{\circ}$ & 10 & 5 & 2 & 5 & 1 & 10 \\ \hline
R7 & 0.05 & $4^{\circ}$ & 10 & 5 & 2 & 5 & 1 & 10 \\ \hline
R8 & 0.10 & $4^{\circ}$ & 10 & 5 & 2 & 5 & 1 & 10 \\ \hline
R9 & 0.05 & $2^{\circ}$ & 10 & 5 & 2 & 4 & 1 & 10 \\ \hline
R10 & 0.05 & $2^{\circ}$ & 10 & 7 & 0.2 & 6 & 1 & 10 \\ \hline
R11 & 0.05 & $2^{\circ}$ & 5 & 3 & 0.2 & 8 & 1 & 10 \\ \hline
\end{tabular}
\end{center}
\caption{\label{tab1} A subset of the 3D models performed. From left to
right, the columns give: class name; mean eccentricity; standard deviation of
inclination; mean mass; standard deviation of mass; lower-mass cutoff;
initial separations in mutual Hill radii; disc mass (normalised against
fiducial value); number of protoplanets.}
\end{table}

Planetary eccentricities were determined by defining a mean eccentricity 
$\mu_e$ for the planetary swarm, and standard deviation $\sigma_e=0.01$, 
with eccentricities then chosen randomly according to a Gaussian distribution. 
Each body is given a random argument of pericentre. Inclinations are likewise
Gaussian distributed according a mean $\mu_i=0^{\circ}$ and standard
deviation $\sigma_i$, with random longitude of ascending node.

Two different schemes were used to determine initial planetary masses. In the 
`ordered' models, the standard procedure was to define the mass of the
innermost body (usually $m_1 = 2$ \me) with subsequent bodies having
$m_{i+1} = m_i + 2$ \me. This somewhat artificial set-up was chosen to maximise
convergent migration, and hence maximise interactions between the bodies.
In the `randomised' models, a more natural distribution is formed from a 
randomly selected Gaussian distribution of planetary masses, with mean $\mu_m$ 
and standard deviation $\sigma_m$, subject to a lower-mass cutoff of either
2 or 0.2 \me, and an upper cutoff of 20 \me~(but note that collisions may
raise masses above this value).

The disc was initialised with scale height $H/r=0.05$ and
$\Sigma(r)=\Sigma_0r^{-0.5}$, where $\Sigma_0$ was typically chosen such that
the disc contains 40 Jupiter masses of gaseous material within 40 AU of the 
central star. Other disc masses and surface density profiles 
were also used, and are described where 
appropriate in subsequent sections of this paper. In the NIRVANA 
simulation the disc had a viscous alpha parameter of $\alpha=5\times10^{-3}$
\citep{shakura73}.

The distribution of semi-major axes, planet and disc masses, and values of 
$\mu_e, \sigma_e$ and $\sigma_i$ together define a class of model. For each 
model class five realisations of the initial data were generated by rotating 
the random number seeds, giving rise to five different simulations. In total
over 300 simulations were run; details of a subset of the models used are 
given in table \ref{tab1}, including those described in subsequent sections.

\begin{figure}
\begin{center}
\includegraphics[width=7.5cm]{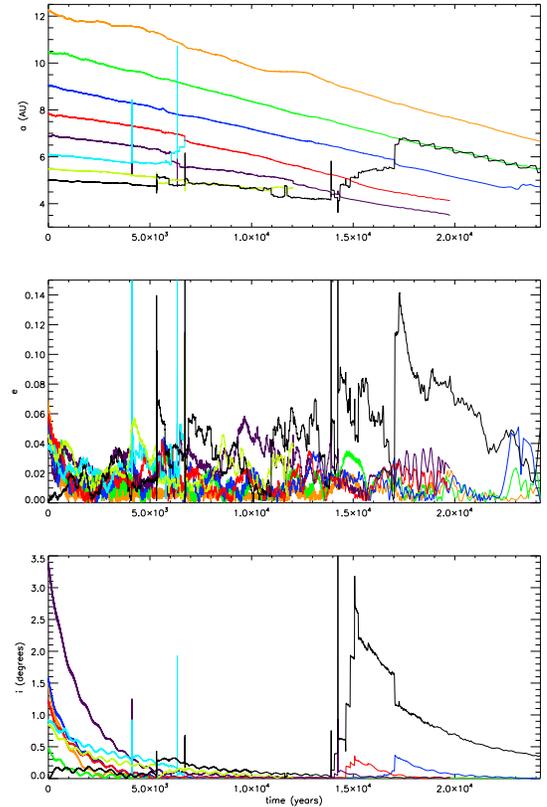}
\caption{\label{fig4} Orbital elements for the 3D NIRVANA model, 
similar to Fig.~\ref{fig5} but with eight protoplanets 
(run H1). Behaviour is 
similar to the early stages of the $N$-body models, with scattering and 
collisions among the inner population. A co-orbital system is formed, which 
remains stable for the remainder of the integrations. Note only the smallest 
protoplanet achieves $e>0.1$ or $i>0.5^{\circ}$ after the initial values have 
been damped.
The 6 \& 10 \me~protoplanets collide at $t=6.7\times10^3$ yrs, 
and the 4 \& 8 \me~protoplanets collide at $t=1.20\times10^4$ yrs.}
\end{center}
\end{figure}

\section{Results of the hydrodynamic simulation}
\label{hydro-results}

One 3D hydrodynamic run (case H1) 
was performed to complement the $N$-body
models. We adopted the fiducial ordered mass set-up, to maximise convergent
differential migration. 
To model the system for over $2\times 10^4$ years while preserving the grid 
resolution, eight protoplanets were used so that the radial extent of the grid 
could be lessened (ten protoplanets were normally used in the $N$-body 
simulations described later). The protoplanets take the same masses as the 
eight inner bodies of the fiducial ordered runs, so that the inner body is 2 
$M_{\oplus}$ and the outer is 16 $M_{\oplus}$. The number of active cells used 
was $(N_r,N_{\theta},N_{\phi})=(288,40,444)$; this provides a resolution finer 
than those used to construct Eqns.~\ref{ecc-eq}--\ref{mig-eq} 
by a factor of 1.5 in azimuth and 2 in radius. As in those tests, for the 
potential softening we adopt a value almost one tenth the vertical height of a 
cell.

The results of this model are shown in Fig.~\ref{fig4}. Resonant migration
dominates the model: after a short period of low-level scattering
and orbital exchanges (where horseshoe interactions rearrange the radial 
ordering of two adjacent protoplanets), the system settled down into two
groups of planets which are in mutual mean-motion resonances, with all 
resonances being either 4:3 or 5:4. The inner pair have suffered collisions 
(at $6.7\times10^3$ and $1.2\times10^4$ yrs)
and end the scattering phase in a 5:4 MMR and on orbits diverging from 
the external protoplanets
due to their increased mass; they are no longer plotted once 
the inner body of the pair enters
the inner damping region, but based on all other models performed, 
we expect this pair would continue to migrate inward in resonance.

Resonant migration is thus the dominant outcome of the simulation.
Only the smallest, scattered body achieves a 
significant eccentricity or inclination, with the other protoplanets limited
to $e<0.1$ and $i<0.5^{\circ}$ once their initial values have been damped
by the disc. By the end of the run all other protoplanet inclinations
are near zero due to the strong damping. In the absence of a
magnetospheric cavity, or other halting
mechanism, we expect all protoplanets to migrate into the
central star.

One further point of note is that a co-orbital pair of planets
are observed. 
Initially this involves the planets undergoing mutual horseshoe motions, 
but it is expected from the results of Paper I that
the damping action of the disc will reduce the libration width until tadpole 
orbits result. The co-orbital pair remains stable in simultaneous resonant 
migration with other bodies on both interior and exterior orbits for the
remainder of the integration.

Despite a total run time of over $1.5\times 10^5$ cpu hours on a 
parallel facility,
due to the high cost of performing hydrodynamic simulations in three 
dimensions we are unable to continue this simulation for more than a few 
$\times 10^4$ years, nor repeat other similar simulations many times in order 
to gain a statistical overview of these chaotic systems. Instead, we perform
numerous $N$-body simulations using the HENC-3D code to examine these issues.

\section{Results of the modified $N$-body simulations}
\label{nbody-results}

We have performed more than 300 modified $N$-body simulations to
examine the evolution of clusters of low-mass protoplanets embedded in a 3D
protoplanetary disc, varying planet masses, orbital parameters, 
and disc masses. Despite
all this variation in initial conditions, 
a number of simple trends are observed. We present a
select few results to illustrate these trends. Further detail of the 2D
analogues of these trends may be found in Paper I.

\subsection{A fiducial ordered mass $N$-body simulation}
\label{ordered-fiducial}

\begin{figure}[t]
\begin{center}
\includegraphics[width=9cm]{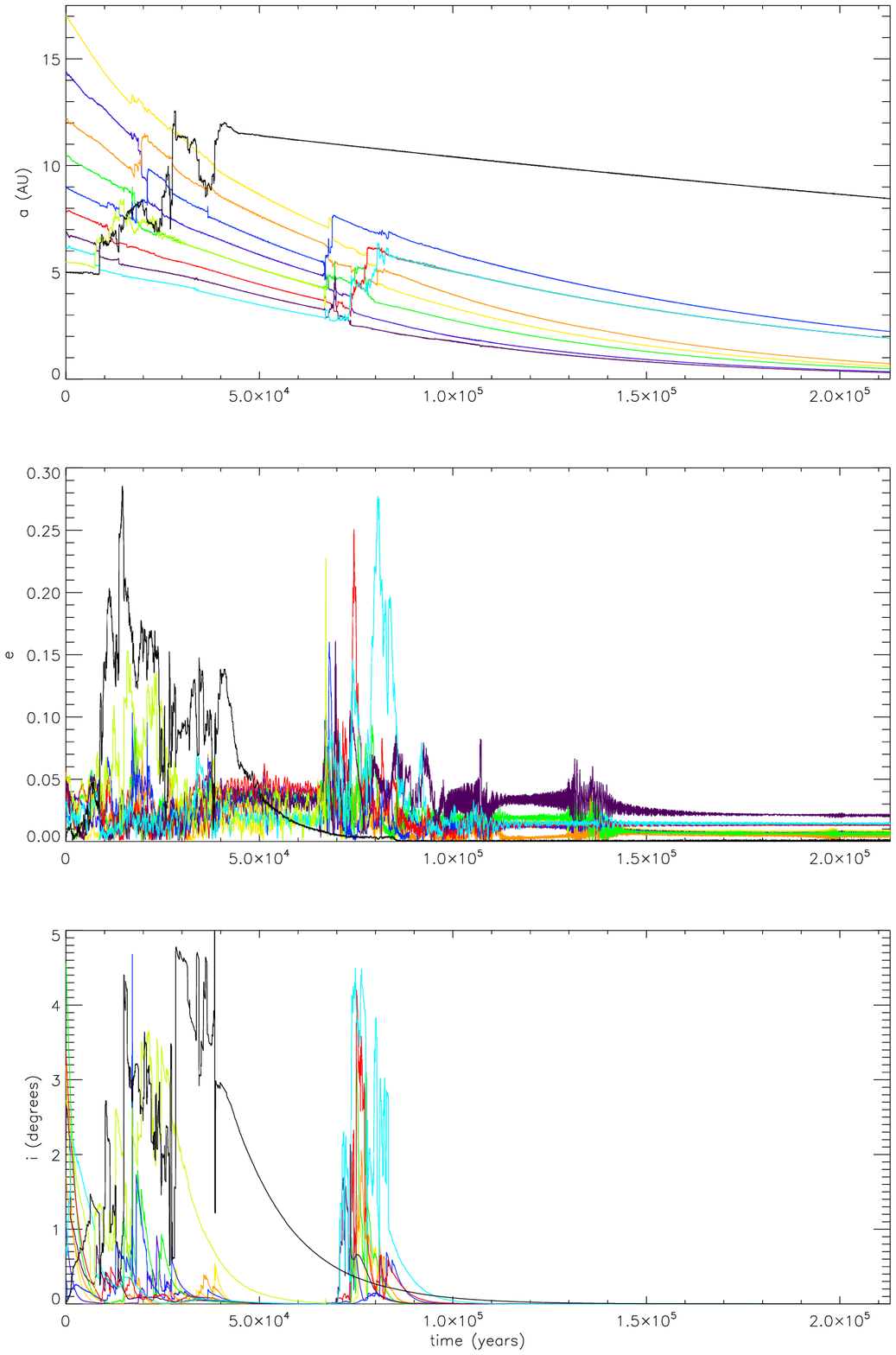}
\caption{\label{fig5} Evolution of a 10 planet $N$-body model with ordered
masses and the fiducial set-up (O1).
({\it Top}) Semi-major axes of the migrating embryos. Short periods of
activity are followed by long periods of migration between bodies in first
order mean-motion resonances.
The 4 \& 18 \me~planets collide at $t=6.76\times10^4$ yrs.
({\it Middle}) The embryos' eccentricities over the same time.
({\it Bottom}) The embryos' inclinations.}
\end{center}
\end{figure}

We choose one particular class of model (class O1 in table \ref{tab1}) 
to act as the fiducial case, against which other models are compared. 
As shown in Fig.~\ref{fig5} the model follows a typical pattern: 
differential migration initially causes the planets to move closer to 
each other. In the outer half of the swarm, orbital exchanges, where two
planets appear to swap semi-major axes in a horseshoe motion, may occur. 
A similar situation is seen among the inner half; however, 
due to the larger mass ratios $m_{i+1}/m_i$ between neighbouring inner bodies
such an exchange
leaves the less massive planet with a slightly larger semi-major axis
than the body it has just replaced, according to the simple ratio
$\Delta a_j/\Delta a_i \simeq m_j / m_i$. The 
smaller body is now closer to the next external mass than the previous planet
was, making further such exchanges (themselves with larger $m_j/m_i$) 
more likely. In this manner the smallest
masses in the population are often passed outward from planet to planet, 
gaining in eccentricity and inclination in the process. This process
typically culminates with: {\it (i)} collision with another body;
{\it (ii)} co-orbital capture; {\it (iii)} ejection beyond the outer edge
of the swarm. A rarer fourth outcome is capture in the Hill sphere of another
body, forming a planet-moon or binary planet system.
HENC-3D is not equipped to model such encounters accurately, which always
result in a collision after a few $\times 10^4$ years.
The process may repeat for several of the smallest bodies,
such that the initially innermost planets finally constitute (in some ordering) 
the outermost planets of the swarm, with the original outermost bodies 
now leading the inward migration having `pushed through' the swarm.

As the population of planets converges due to differential migration,
a series of mean-motion resonances (MMRs) 
forms between the planets (or subsets of them),
forcing them to migrate inward in lockstep.
Occasional bursts of dynamical instability occur, leading to the
removal of bodies by collision, co-orbital capture or ejection to the outer
edge of the swarm, and this helps stabilise the population.
The initial phase of scattering smaller bodies ends 
within a few thousand years, 
and the disc rapidly damps eccentricies and inclinations developed during
this active phase. During the ensuing resonant migration, as separations 
between the planets decrease,
a single planet is occasionally perturbed sufficiently to break the resonant 
chain, and the process of the swarm reordering itself
may repeat on a shorter time scale (often $<10^3$ yrs), as seen in 
Fig.~\ref{fig5} at time $t \approx 6.5\times10^5$ years.
Ultimately the population, now comprised of
an individual planet and two resonant groups, 
migrates towards the central star, where it will be accreted in 
the absence of a 
stopping mechanism, such as a 
disc edge created by a magnetospheric cavity,
or a `planet trap' associated with a region of
the disc with a positive surface density gradient \citep{masset06}.
The final masses of the planets, labelled from smaller to larger radii at
the end of the simulation, are (first group) $8,22,14,20,16$ \me, 
(second group, including the co-orbital pair) $10,6,12$ \me, 
and $2$ \me, showing how the 
population has separated with the smallest bodies at larger radii and the 
largest bodies closer to the star.
The smallest planets, if scattered significantly far
outwards, or with sufficient eccentricity or inclination to slow their
migration, may survive for times on the order of $10^6$ years, raising the 
possibility of survival as a super-terrestrial body if the 
population formed late in the disc's lifetime.

\begin{figure*}[t]
\begin{center}
\includegraphics[width=16cm]{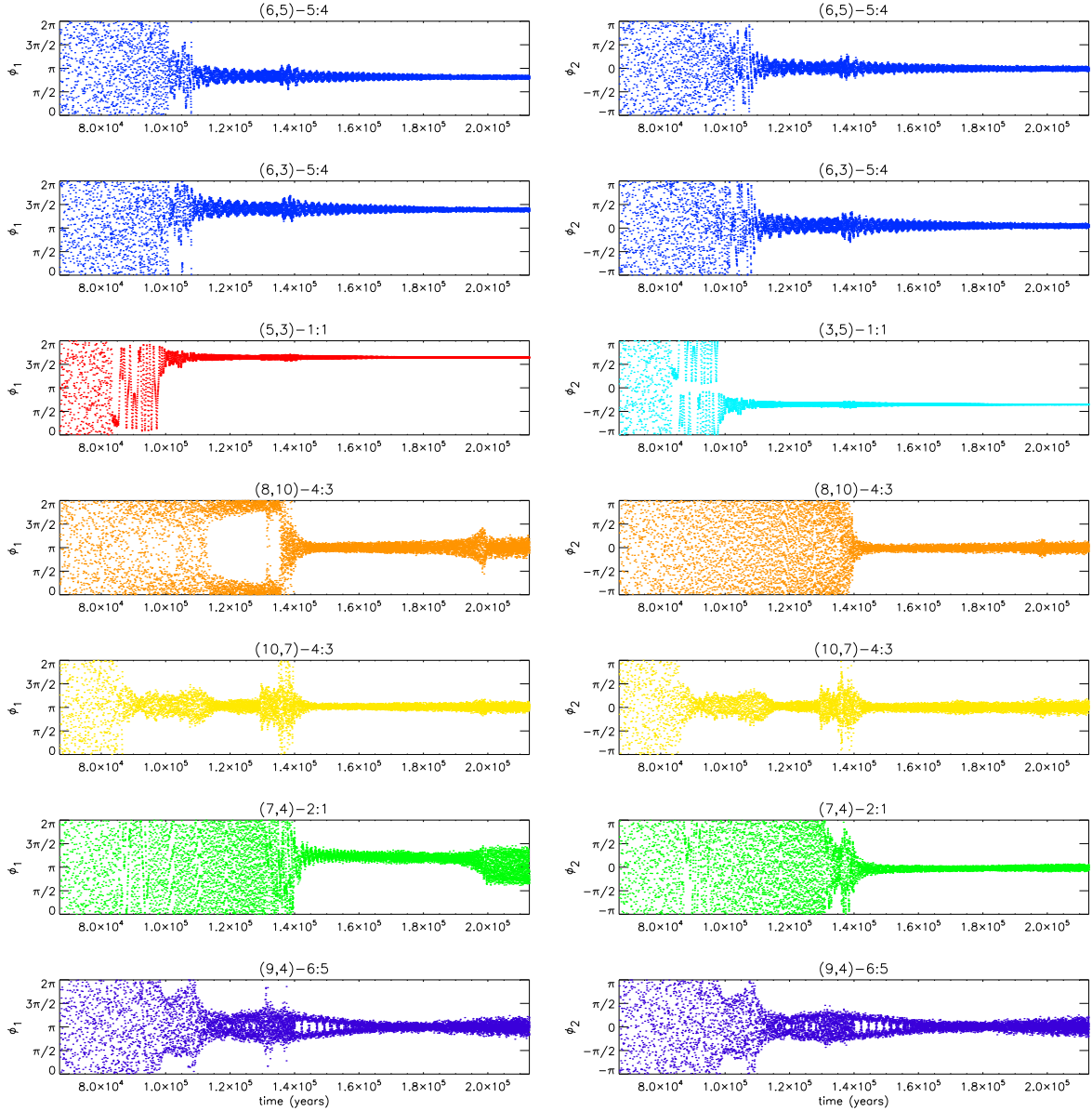}
\end{center}
\caption{\label{fig6} The first order mean-motion resonances between the 
planets in Fig.~\ref{fig5}, during the longest resonant migration phase.
Each plot is labelled (P$_i$,P$_j$)--$p$:$q$, where
P$_i$ and P$_j$ are the exterior and interior protoplanets in each resonance
respectively, and $p$:$q$ is the commensurability between them.}
\end{figure*}

Due to the strong inclination damping, the system is approximately planar
for most of its duration. 
As the planets migrate together, the MMRs between neighbouring bodies are 
almost always ($> 99$\%) first order, typically 3:2, 4:3,\ldots,8:7. The
first order and co-orbital resonances between the planets in Fig.~\ref{fig5} 
are shown in Fig.~\ref{fig6}, where the planets are labelled 
1--10 with 1 being the initially innermost body, and 10 the initially 
outermost. The resonant angles for the first order resonances ($p=q+1$) 
are defined by:
\begin{eqnarray}
\label{resonance-eq}
\phi_1 & = & p\lambda_1 - q\lambda_2 - \omega_1 \nonumber \\ 
\phi_2 & = & p\lambda_1 - q\lambda_2 - \omega_2 
\end{eqnarray}
where $\lambda_1$ ($\lambda_2$) and $\omega_1$ ($\omega_2$) 
are the mean longitude and 
longitude of pericentre for the outer (inner) planet, respectively. 
Commonly, the complicated lattice of resonances between groups of planets 
generates additional first, second and third order resonances between 
non-neighbouring planets ({\it e.g.}~ alternating 5:4 and 6:5 resonances 
implies the existence of a 3:2 resonance between the non-neighbouring planets).
Rather than being a source of perturbation to resonant pairs, additional
planets thus often provide further stability to a migrating resonant group.
 
In Sect.~\ref{statistics} we present a more statistical 
analysis of our results,
but in the next subsection we discuss how the qualitative nature of
the results change as initial conditions are varied.

\subsection{Variation of the initial conditions}
\label{ordered-vary}

The behaviour discussed above represents the typical behaviour observed among 
the ordered mass models. Many other sets of initial conditions were used, 
including varying initial eccentricities and inclinations up to 
$\mu_e=0.15$ and $\sigma_i=6^{\circ}$ ({\it e.g.~} runs O2-O4). 

\subsubsection{Variation of initial eccentricities}
Increasing the initial eccentricities typically produces a much longer
dynamically active phase among the swarm, which
involves more planets and prevents the formation of early but
temporary resonant groups. This is especially true when the initial
eccentricities are such that protoplanets begin on crossing orbits. 
Much of the population is involved in dynamical relaxation until
several bodies have been expelled from the group, or lost in collisions.
In the long run, however, the final states of these systems are
similar, and consist of inwardly migrating groups of planets, in multiple
mean-motion resonances, often including co-orbital systems. Their final
fate remains accretion by the central star.

\subsubsection{Variation of initial inclinations}
Raising initial inclinations has little effect, as they
are damped by the disc before large-scale scattering activity begins.
This is to be expected as raising inclinations does not lead to the creation
of crossing orbits, unlike the initial eccentricity variations. Systems
with large values of initial inclination have their inclinations damped
on the time scale of a few thousand orbits, after which they
enter a phase of dynamical interaction similar to that shown by the fiducial
run described in Sect.~\ref{ordered-fiducial}.

Scattering throughout the population is now more common overall than in 2D 
(see Sect.~\ref{statistics}) and the resulting activity largely removes
memory of the initial conditions.
For the initial eccentricity/inclination distribution to substantially
alter the long term evolution requires seemingly unphysical values
for these quantities, corresponding to non-disc-like initial conditions.

\subsubsection{Variation of initial separations}
The initial separations between the planets were varied, with values
ranging between 4 and 8 $R_{\rm{mH}}$.
This was found to have no qualitative effect on the final
states of the systems. Early stages showed significant differences, with
closer protoplanets undergoing prolonged periods of
violent excitation, and more distant ones slowly
moving closer together and taking longer to produce any significant 
scattering activity. Once scattering activity ceased to continue
the final states of the systems were again very similar to those already described.

\subsubsection{Modifying the disc mass}
Altering the disc mass produced more substantial changes. Reduction by a factor
of 2 (class O7) produced population-wide gravitational interactions for similar 
lengths of time ($\sim 1$--$2 \times 10^5$ years) despite longer migration 
times, yet the swarm was more likely to retain its original order, with fewer 
of the more massive planets passing through the population and instead 
remaining at the rear of the population where they drive the whole group 
forward. This presumably arises because of the overall
reduction in convergent migration rates. 
The resonant groups were often larger, comparable to 
those seen in 2D (consisting of up to 7 bodies). 
However, those small planets that 
were scattered tended to achieve higher eccentricities ($e \simeq 0.5$--0.6) 
and higher 
inclinations (6$^\circ$--12$^\circ$), and remained in such excited orbits for 
several times longer than in the fiducial case. By spending significant 
portions of each of orbit away from the other protoplanets
(due to the high inclinations),
encounters that may result in collisions or co-orbital pairs 
become less likely, allowing the smaller body to retain its 
excited state for longer. 
Further reductions in disc mass increased the effect 
slightly and were more likely to produce larger resonant groups. 

\subsubsection{Changing the number of planets}
We have run models with smaller numbers of protoplanets (five rather than ten).
One such model is listed as class O9 in table \ref{tab1}.
Although the early phase of evolution can occasionally involve
all planets being in mutual mean-motion resonances, this
configuration was found to be stable for long time scales in 
only approx.~20\% of runs; 
interestingly, only those models that formed a stable co-orbital pair
were able to sustain such a five-planet resonant group.
In all other cases instabilities leading to scattering reduced the group size, 
either by ejecting one of the smallest protoplanets to a larger orbit (most common
when the initial mass range of protoplanets was largest, 2--20 \me) or by 
removing one or more protoplanets by collision (most common when the initial mass 
range was smallest, 2--10 \me).
In all cases the long-term evolution always resulted in migration of the 
remaining protoplanetary swarm into the central star.

\subsection{A fiducial randomised mass $N$-body simulation}
\label{random-fiducial}

\begin{figure}[t]
\begin{center}
\includegraphics[width=9cm]{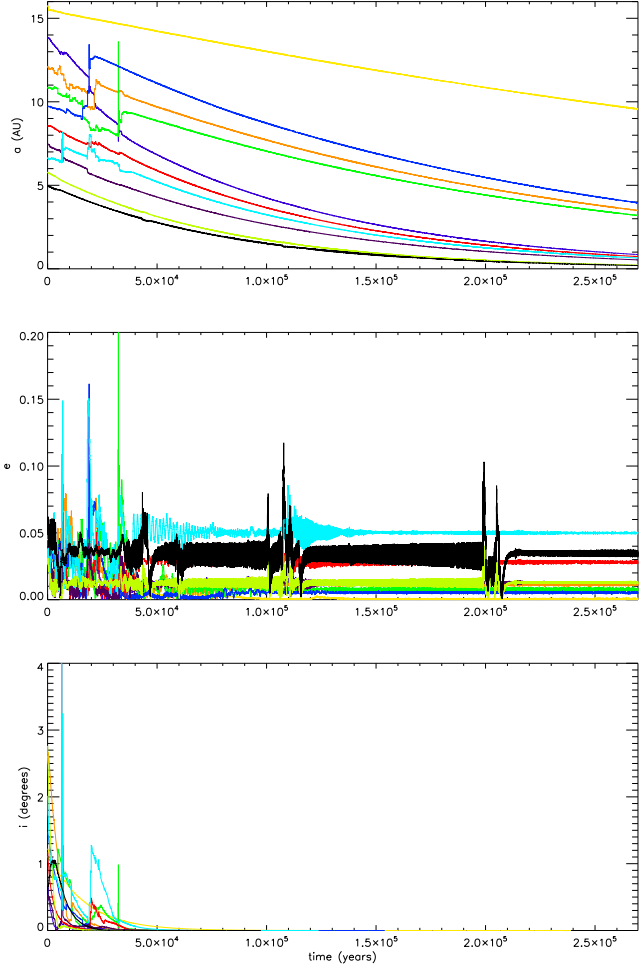}
\caption{\label{fig7} Evolution of a 10 planet $N$-body model with randomised
masses and the fiducial set-up (R1).
({\it Top}) Semi-major axes of the migrating embryos. Short periods of
activity are followed by long periods of migration of bodies in first
order mean-motion resonances.
({\it Middle}) The embryos' eccentricities over the same time.
({\it Bottom}) The embryos' inclinations.}
\end{center}
\end{figure}
When initiating 
models with randomised masses, we set a mean $\mu_m$ and standard deviation
$\sigma_m$ for the mass distribution, and chose masses randomly
according to a Gaussian distribution. A wide
range of mass distributions were considered, which were
conflated with the variations in initial
conditions used in the ordered mass models. For a fiducial randomised model,
we choose initial conditions and the planetary mass range
similar to the fiducial ordered case (class R1). Such a
model is shown in Fig.~\ref{fig7}.

\begin{figure}[t]
\begin{center}
\includegraphics[width=9cm]{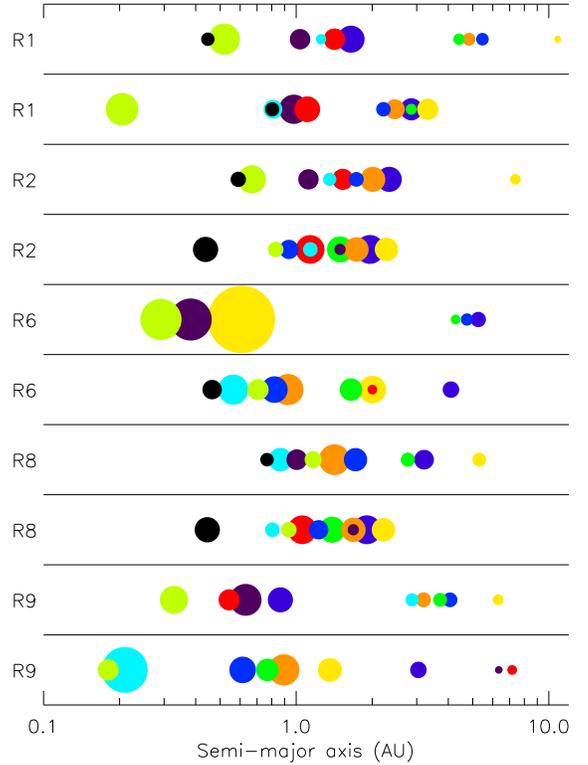}
\caption{\label{fig8}Snapshots of two randomly selected simulations from the
model classes shown, with the exception of the first (R1) and fourth (R2) rows,
which correspond to Figs.~\ref{fig7} \& \ref{fig9} respectively. The snapshots 
are taken at $t=2\times10^5$ yrs. The area of each planet is linearly 
proportional to its mass; the smallest body (in the top row) 
is of 2 \me, and the largest (formed from multiple collisions) is of 39 \me.
The co-orbitals of Fig.~\ref{fig9} are clearly visible as concentric
circles. }
\end{center}
\end{figure}

Overall, randomised models follow broadly similar evolutionary paths 
to the ordered models, except that differential migration 
carries some planets, or groups of planets, away from each other.
Consequently, unless the early gravitational
interactions among the population are prolonged or involve especially
strong scattering, the
protoplanets rapidly ($< 5 \times 10^4$ yrs) break up into smaller groups 
of typically 2--5 bodies,  with the slowest 
migrators typically following behind in isolation after scattering. This 
behaviour is clearly seen in Fig.~\ref{fig7} 
and also in Fig.~\ref{fig8}, 
which shows snapshots of the end-states of a random selection of models. 
Planets of the largest or smallest mass typically end the simulation migrating
alone or in small groups at the inner or outer edges of the population,
respectively.

\subsubsection{Variation of the mass distribution}
\label{random-vary}

Several different mass distributions were combined
with the various initial eccentricity and inclination
distributions that we have considered previously.
Values in the range 5 \me~$< \mu_m < 10$ \me~and
3 \me~$< \sigma_m < 7$ \me~({\it e.g.~} classes R1--R5) were
used in setting up initial protoplanetary swarms.
The resulting  models can be characterised by a number of features.

First it was found that among the primary mass ranges studied 
($5 \leq \mu_m \leq 10$ \me), when $\sigma_m \la \mu_m/3$ 
scattering was substantially reduced, resulting typically in
one large group migrating without further strong interaction
after an initial period of orbital readjustment. This is clearly
due to the fact that the planets are migrating inward at similar
rates. For larger
ratios of $\sigma_m/\mu_m$ 
planet-planet scattering and its natural consequences 
(collisions, co-orbital
planets, {\it etc.}) were more common, due to the concomitant
increase in differential migration.
Once strong scattering was induced, however, we found no strong
correlations between the final outcomes and the value of $\sigma_m/\mu_m$. 

The second feature we note is that additional tests for lower-mass 
planets (see Sect.~\ref{limits} for a more complete discussion), using
mean masses in the interval ($0.3 \la \mu_m \la 2$ \me), seem to
result in more energetic and population-wide
scattering activity than is the case for higher-mass planets,
which may seem surprising at first glance.
This result appears to originate in the fact that
the lower-mass planets tend to become trapped in first order
resonances of high degree (even as large as 14:13 for the lowest
mass cases considered), placing the
planets in very close proximity. Localised instabilites in the resonantly
migrating swarm then lead to strong dynamical interaction and scattering.
Such large-scale activity was also seen among ordered mass 
runs using similarly small initial masses.

The third feature shown by the randomised mass runs is that
the initial distribution in semi-major axes of the bodies
determines much of the subsequent evolution.
If the innermost bodies are of a higher mass than the mean of the
population, those planets will form a separate group and migrate 
away from the rest, generally leading to an uneventful inward
journey for the whole population. 
Conversely, if several massive bodies are located in the outer half of
the population, then they are likely to
drive a larger resonant group ahead of
themselves, usually causing the smallest bodies to be scattered outwards,
and also more likely to generate co-orbital systems. 

The fourth feature we observe is that the initial 
distribution of masses has more effect on the
resulting level of activity than their initial separations. That is, although
it may take longer for differential migration to bring together initially 
widely separated bodies, the resulting activity is determined by the mass
distribution just as it is when the protoplanets are initially closer together.
Separations of up to 20 \rmh~have been tested, with the mass distribution
remaining the dominant factor. 

Together, these features suggests that the level
of activity among a population of moderate to large cores,
assuming a moderate spacing of several mutual Hill radii, is dependent on the 
mass ratios of nearby bodies, with little scattering 
activity below a critical value 
$\sigma_m/\mu_m \la 1/3$ for a Gaussian distribution. Collections of smaller
masses appear to undergo periods of evolution where the degree
of scattering activity is greater than for their more massive
counterparts.
We note that the smaller protoplanets generally form MMRs of
higher degree, in the range 8:7--10:9 (or in more extreme cases 14:13), 
which
in conjunction with the weaker damping forces on eccentricity
and inclination
may account for the increased scattering due to closer proximity.
Nonetheless, we observed that in some cases these resonances
could remain stable over long periods, and examples of long term
stable planets in the 14:13 resonance occurred in some simulations.

Varying the mass distribution produced no significant correlation with any 
other parameter variation (Sect.~\ref{ordered-vary}).

\subsection{Prevalence of co-orbital systems}
\label{co-orbitals}

A significant feature that arose in Paper I was the unexpected abundance of
stable co-orbital planets, orbiting around their mutual $L_4/L_5$ points. 
Co-orbital systems form when planet-planet scattering causes a planet to
be perturbed such that its semi-major axis becomes very similar
to that of another planet in the system. Disc-induced eccentricity damping
then ensures rapid decay of the planet eccentricity, leading to
co-orbital capture. Within the planetary swarm, the condition for long-term
capture is simply that this eccentricity decay occurs before the
scattered planet undergoes a close encounter with another protoplanet, which
would otherwise disrupt the co-orbital system.
We find that many of the co-orbital systems 
remain stable for the duration of the 
simulation while migrating inward over distances greater than 10 AU.

In 3D, we find that these co-orbital planets
are more common than in 2D (see Fig.~\ref{fig10}), 
occurring in almost 45\% of ordered and almost 35\% of 
randomised simulations; of these, approximately 21\% and 14\% respectively 
contained multiple examples of stable,
co-orbital pairs in the same simulation
(as with other resonances, co-orbital pairs are deemed stable within a 
simulation if they survive for a time  $>10^5$ years.) 
We note that the eccentricity damping rate adopted in the 2D simulations
of Paper I was smaller than that used for the 3D runs in this paper.
We attribute this increased co-orbital frequency to the 
higher eccentricity damping rate and the generally smaller sizes of the 
resonant groups ---
both of which reduce the opportunity for a recently captured co-orbital body
to be disturbed by other protoplanets --- together with a slight increase in 
overall scattering activity (Sect.~\ref{statistics}).
Inclined orbits are not an obstacle to co-orbital
formation, and planets with inclinations greater
than $10^{\circ}$ are readily captured
into stable horseshoe orbits, with these mutual inclinations 
eventually being damped by the disc. 

When a co-orbital system first forms it is usually in
a mutual horsehoe orbit.
In all except one instance the horseshoe motions decayed because of the disc's
action to form tadpole orbits, maintaining small oscillations about the 
$L_4$/$L_5$ points. The transition from horseshoe to tadpole motion typically 
takes $0.3$--1$\times 10^4$ years for our standard disc parameters. 

\begin{figure}[!t]
\begin{center}
\includegraphics[width=6cm]{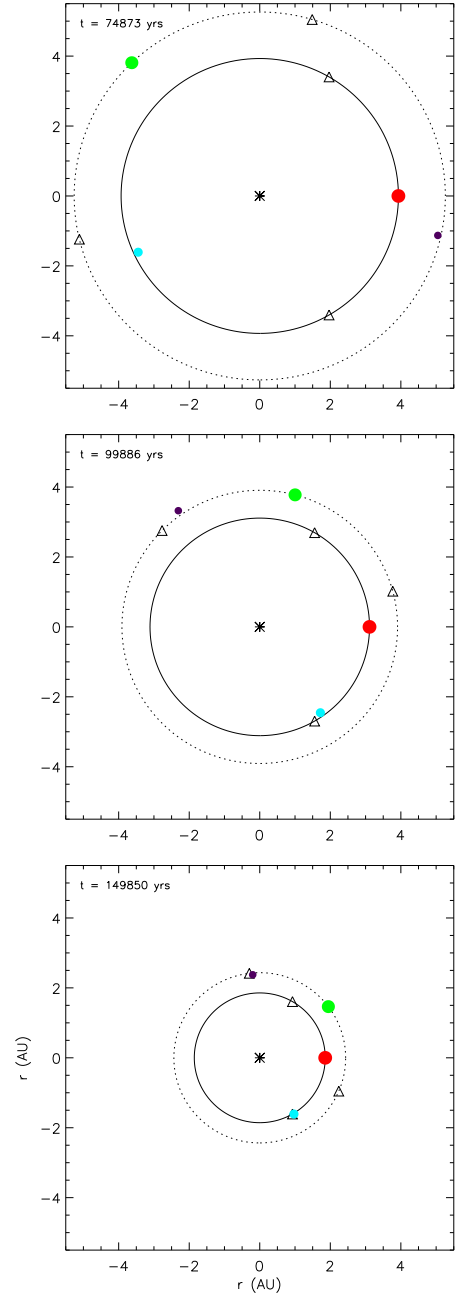}
\caption{\label{fig9} Evolution of a co-orbital system (class R2)
migrating inwards: two co-orbital pairs, locked in a 3:2 MMR. 
The size of each planet is proportional to
its mass, while the open triangles display each primary body's $L_4$ \& 
$L_5$ points. The other protoplanets present in the model are not displayed, 
but those shown are part of a resonant group encompassing nine bodies.
({\it Top}) Initially the co-orbital planets' motions cover the whole horseshoe region.
({\it Middle}) The disc's action causes the planets to shift into tadpole orbits.
({\it Bottom}) The planets migrate inwards under small librations.}
\end{center}
\end{figure}

A detail of two migrating co-orbital systems (from class R2)
is shown in Fig.~\ref{fig9}. 
Initially the horseshoe librations are of large amplitude, 
but these decrease with time, and
after 2--3 $\times 10^3$ yrs tadpole motions result.
The system then contines to migrate inward maintaining this architecture.

Co-orbital planets are found both as isolated pairs, and as part of
larger resonant groups, sharing mean-motion resonances with external and 
internal bodies like any other individual body. 
At the end-state of a simulation, a co-orbital pair may thus
be in multiple first- and second-order MMRs with two or more interior/exterior 
bodies, all densely packed within 0.4 AU of the star.

\subsubsection{Limits on co-orbital formation}
\label{limits}

Other studies have examined the behaviour of swarms of protoplanets,
subject to type I migration, at earlier stages of formation when
protoplanetary masses are considerably smaller ({\it e.g.~}\citealt{mcneil05}).
Similar patterns of resonant migration have been observed, but co-orbital
planets were not found, in contrast to their ubiquity here 
and in Paper I.
Consequently, we seek to determine limits on co-orbital formation and
explain this discrepancy between models; we focus on planetary mass,
initial separations, and the surface density profile of the disc.

To compare with studies conducted using lower-mass protoplanetary swarms,
we selected one fiducial class from each of the ordered (O1) and randomised (R1)
models, and reran them with the protoplanets' masses 
($\mu_m$ and $\sigma_m$, in the randomised case)
repeatedly halved; the lower-mass cut-offs are set to 0
in both cases (see Sect.~\ref{random-vary}). 
Additionally, in the randomised models we increased the initial
separations in stages up to 20 \rmh.

When using our standard disc model with these smaller protoplanet masses,
25\% of simulations resulted in stable
co-orbital pairs for both the ordered and randomised
mass distributions, averaged over the ranges of planetary
masses and initial separations considered. 

With smaller sample sizes, we can be less confident about these statistics
than for the main body of the simulations performed. However, the fact that
co-orbitals continue to appear at all, and with frequencies of $\sim 25$\%, 
is significant. The region within which co-orbital capture can occur
scales $\sim m_p^{1/3}$, and so we may expect capture rates to fall to near
zero for populations of such small bodies. However, differential migration
brings protoplanets closer together; as masses decrease, the MMRs formed
between neighbouring protoplanets are of higher degree (stable resonances up to
14:13 were observed among this work, as mentioned previously) 
and so the planets lie closer together.
Consequently only a small shift in the orbit is necessary for one body to enter
the horseshoe region of another and be captured, and a perturbation from a
third body can provide the necessary energy to jump into the horseshoe region.
All that is required for co-orbital formation is 3--4 protoplanets in
close proximity.

We also note that, for the randomised mass distributions, co-orbital frequency
is essentially independent of the initial separations. This is easily
interpreted as resonant migration driving groups of bodies together;
differential migration may cause a relatively massive protoplanet to `sweep up'
several smaller bodies in MMRs ahead of it. Formation of a co-orbital pair from
such collections is then dependent only on the usual mutual interactions within
a small group. Provided the initial separations are not so large that such
groups do not have time to form --- highly unlikely for most oligarchic
growth scenarios, {\it e.g.}~\citet{kokubo00} --- co-orbital formation thus
remains primarily dependent on the initial mass distribution.

One feature different here from the set-up of \citet{mcneil05} are the
disc damping times, where they utilised the higher damping rates of
\citet{papaloizou00}; the discrepancy is approximately given by a factor of 3
(see also Figs.~1 \& 2 of Paper I).
To test the effect of this difference, we shortened the damping time in
Eqn.~\ref{ecc-eq} by this factor, and reran a selection of models with the
lowest mass distributions previously considered. We find that although
interactions between the planets are significantly reduced, 
co-orbital planets are
still able to form as differential migration brings a population together, with
crowding producing the minor perturbations and individual orbital exchanges
required, despite low eccentricities among the group.

The situation changes dramatically when the disc surface
density profile is steepened. We ran a suite of models
with the power law exponent being decreased to $-3/2$,
which is the value usually adopted for the minimum mass
solar nebula \citep{hayashi81}, and corresponds to the disc
model used in McNeil et al (2005).  
In this case the incidence of co-orbital system formation
fell to almost zero, this being a direct consequence of the fact
that the migration of interior bodies is speeded up,
and that of bodies lying further out in the disc is
slowed down, when the density profile is steepened. 
Thus, it would appear that co-orbital planet formation
is highly sensitive to there being strong
differential migration that brings bodies together,
a situation that is highly favoured in discs with flatter
surface density profiles.
Observation of co-orbital extrasolar planets would thus
be an indicator of strong dynamical relaxation occuring during
planet formation, induced and maintained in-part by there having
been a suitably flat radial surface density profile in the original
protoplanetary disc.

\subsubsection{Long-term stability}
\label{stability}

To test the stability of these tadpole planets during and
beyond disc dispersal, 
a time-dependent function was added to Eq.~\ref{twave-eq} that 
reduced the disc-induced forces acting on the planets
exponentially, simulating
disc mass loss according to a prescribed time scale. We then selected several
simulations that produced tadpole planets, removed the other
planets, and restarted the model after the formation of the 
co-orbital pair. Each pair was found to separate if the disc forces were
reduced too rapidly, typically corresponding to a (grossly unphysical) 
mass-halving time of under $10^3$ years. In all other cases, the co-orbital 
pair remained stable, with their orbits remaining largely unchanged 
once the remaining simulated disc mass became negligible.

Test calculations were also run in which the non-co-orbital planets were
not removed from the system, and these contained instances where:
the co-orbital pair was largely isolated from all other bodies because
of prior differential migration;
the co-orbital pair were on significantly eccentric orbits 
(both planets possessing $e>0.25$); 
the co-orbital pair were in resonance with additional 
bodies driving faster inward migration 
(while the migration force remained effective).
Subject to the condition on disc mass-halving time stated above,
all these models were found to be stable for
as long as the integrations were continued, with a minimum simulation time
of $\sim 2 \times 10^5$ years in each case and a typical duration an order of
magnitude greater.
One model was allowed to evolve in a disc-free environment 
for over $4 \times 10^9$  years, with the co-orbital system remaining stable
for this time.

We note those models with additional planets in mean-motion resonance
with the co-orbital pair also implies the long term stability of pairs of 
protoplanets in MMR after disc dispersal, and these resonant
systems often contain numerous bodies in mutual MMRs. 
While the observed exoplanets
in MMR are giant planets ({\it e.g.~}GJ 876, 55 Cancri, HD 128311, HD 82943, 
HD 7352), 
whose resonances are thought to have been established through differential
type II migration ({\it e.g.~}\citealt{snellgrove01, lee02, kley04}), 
our simulations show that differential type I
migration may lead to the formation of multiple MMRs between groups of 
lower-mass planets. These will 
become amenable to detection as observational techniques
allow greater exploration of the lower-mass end of the
extrasolar planet population.

\subsection{Statistics of simulation outcomes}
\label{statistics}
\begin{figure}[t]
\begin{center}
\includegraphics[width=7cm]{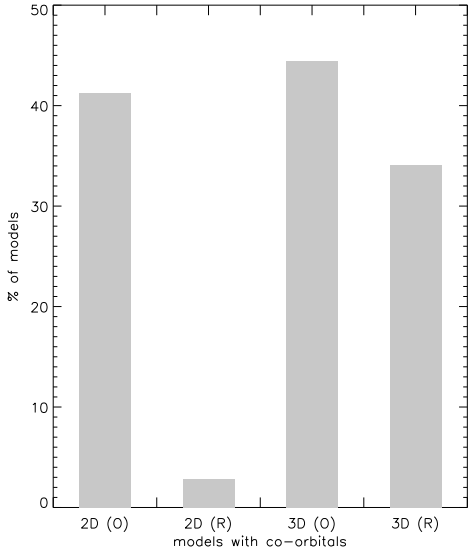}
\caption{\label{fig10} Frequency of $N$-body models displaying stable 
horseshoe or tadpole planets. 
On the left, $N$-body data from Paper I is shown for comparison. 
Left--right: 2D ordered, 2D randomised, 3D ordered, 3D randomised.}
\end{center}
\end{figure}

We now discuss the frequency with which certain simulation outcomes,
such as planet-planet collisions, co-orbital system formation, {\it etc.~}arose.
Figure \ref{fig10} displays the
frequency with which co-orbital systems formed within the simulations,
and survived for at least $10^5$ years. The two leftmost bars
show results for the 2D simulations presented in Paper I.
The rightmost bars show results for the 3D runs presented
in this paper. In the cases where the protoplanet mass increased
with increasing initial semi-major axis,
we see that the 2D and 3D results are very
similar, with $\sim 40$\% of each set of calculations producing long-lived
co-orbital systems. The situation for randomised mass distributions
is different, however, with only 3\% of 2D runs leading to
co-orbital formation, whereas 34\% of 3D runs resulted in long-term
co-orbital formation 
We believe this is due to a stronger eccentricity damping rate (in line
with \citealt{tanaka04}) in the 3D models, which allows the orbits of recently 
captured co-orbital bodies to be circularised and achieve a tadpole orbit 
sooner, reducing the probability of the body encountering a third protoplanet 
and being scattered from its horeshoe orbit. The effect is lessened among
the models with initially ordered mass distributions because these favour
large groups of protoplanets, meaning such
a scattered body is likely to encounter further protoplanets and have multiple
opportunities to be captured as a co-orbital entity.

\begin{figure}[!h]
\begin{center}
\includegraphics[width=7cm]{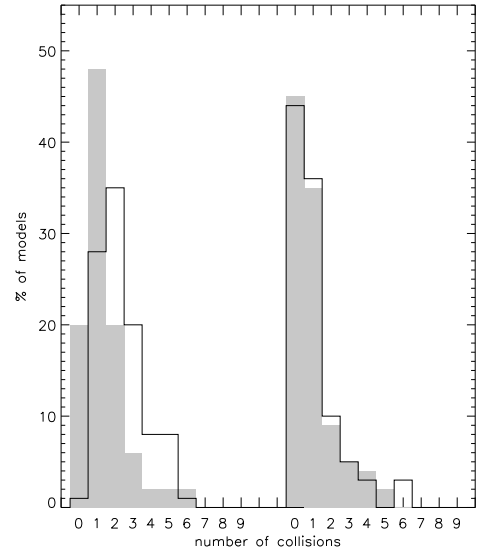}
\caption{\label{fig11} Frequency of collisions per simulation for the 
ordered mass models (left) and randomised mass models (right).
The open bars show the corresponding 2D data of Paper I.}
\end{center}
\end{figure}

Figure \ref{fig11} shows the collision frequency for
2D and 3D simulations, as a function of the number of collisions
occurring per simulation. The left hand charts show
results for runs where the initial protoplanet mass increased with
semi-major axis, the right hand charts show results for
randomised mass distributions. Although there are small
differences between both 2D and 3D runs, and between ordered and
randomised mass distributions, the striking feature of these
plots is their similarity. One normally expects that 2D and
3D simulations would produce different collision frequencies
(with 2D runs leading to many more collisions), but this is not
borne out by our results. The reason is most likely to be that
the very strong inclincation damping provided by the disc causes the
planetary swarms to remain quasi-2D, thus increasing the
collision frequency. 

\begin{figure}[t]
\begin{center}
\includegraphics[width=7cm]{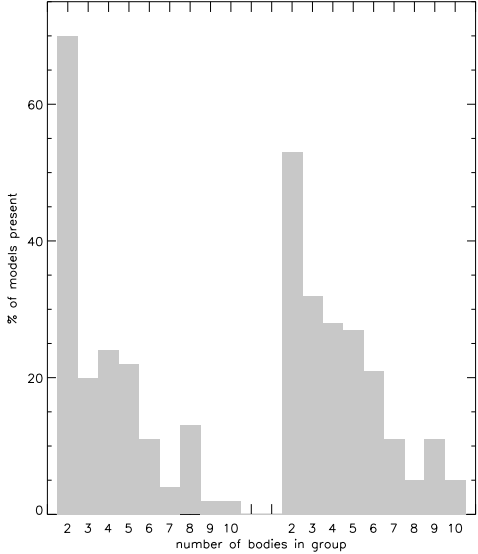}
\caption{\label{fig12} Frequency of resonant group size per simulation for 
the ordered mass models (left) and randomised mass models (right).} 
\end{center}
\end{figure}

Figure \ref{fig12} shows the frequency with which resonant groups
form as a function of the number of protoplanets contained in the resonant
group for 3D simulations only. Resonant groups are only counted in simulations if
they survive for $10^5$ years or more. The left hand chart is
for models in which the initial planet mass increased with
semi-major axis, and that on the right hand side is for the randomised
mass distributions. 
The 3D runs show that simulations containing resonant
groups of just two planets are most
common, but with a significant number containing three to six planets.
This is a clear indication that resonances involving smaller numbers of 
planets are more stable over the longer term.

\section{Conclusions}
\label{conclusions}

We have performed simulations, using two different numerical schemes, 
which examine the 
evolution of swarms of low-mass protoplanets embedded in a 3D 
protoplanetary disc. 
First, we used
hydrodynamical simulations to model the orbital
evolution of planets on initially eccentric and/or inclined orbits,
and fitted analytic equations for the rate of
eccentricity and inclination damping, and migration, to these simulation
results. These equations were then incorporated into an $N$-body
code, which was used to perform many simulations of
planetary swarms embedded in protoplanetary discs.
We also performed a single hydrodynamic simulation of
a planetary swarm embedded in a disc to verify the qualitative
outcome of the $N$-body simulations. Although this simulation
could not be evolved for as long as the $N$-body runs due to computational cost,
we found basic agreement between it and the $N$-body simulations for the
first $\simeq 30,000 $ years of evolution.

The main aim of this work was to re-examine previous results
we had obtained using 2D simulations (see Paper I).
It is known that type I migration of low-mass protoplanets
may be slowed or even reversed by
sustaining significant eccentricities \citep{papaloizou00},
and in Paper I we examined whether gravitational interactions between 
a swarm of protoplanets can provide
the necessary excitation of eccentricitity to prevent inward migration.
The conclusion of that work was that disc-induced eccentricity damping is
too strong, and ultimately protoplanetary swarms migrate into the
central star.
In this paper we have examined whether the inclusion of 3D effects
changes this basic conclusion, as 
the dynamics in 3D can be quite different. For example, the
collision frequency may be reduced, and the possible
excitation of significant inclination may assist in slowing
down migration.

Our results indicate that the collision 
rate is reduced only marginally in 3D, because the strong disc-induced
inclination damping causes the system to remain quasi-2D.
We find that the disc-induced eccentricity damping remains too strong,
so that a protoplanetary swarm is unable to maintain long epochs of
strong gravitational scattering with concomitant high eccentricities.

Despite a wide variety of initial conditions, a typical mode of behaviour
is observed 
across the models, which may be summarised as follows: 
Differential type I migration
leads to converging orbits and a crowded system. Gravitational scattering leads
to orbital exchanges between neighbouring planets, formation of
groups of planets in mutual mean-motion resonances,
formation of 1:1 co-orbital
resonances, and occasional collisions. Smaller protoplanets are
frequently scattered out beyond the population, 
but rarely achieve a semi-major axis greater than that of the 
initially outermost protoplanet prior to migration, nor an eccentricity
or inclination capable of significantly prolonging their life times. 
Within $< 10^5$ yrs of commencement, the simulated
system has typically settled down into 
a state with the protoplanets being in resonant groups,  with each member
of the group being in a mutual mean-motion resonance.
These are usually first 
order mean-motion resonances, typically of degree  3:2--8:7. 
These resonant groups 
migrate inward in lockstep, to be accreted by the central star in the absence
of a stopping mechanism such as an inner magnetospheric cavity.
Occasionally a late burst of scattering activity occurs once 
a swarm of four or more
bodies has migrated over several AU, but these systems always settle down to
another phase of resonant migration that takes them to the star.
A small, slowly migrating protoplanet may
sometimes ($\approx 1$\% of runs) be 
scattered to the outer edge of the swarm (where it cannot be trapped in 
resonance by a faster migrating body) and survive for $\sim 10^6$ yrs. 

We conclude that
if multiple protoplanets form coevally from oligarchic growth in the giant 
planet zone of a laminar disc, then the long term evolution of the system will
usually be collective inward migration and ultimate accretion by 
the central star.
This occurs on a time scale much shorter than the accretion time required
to accrete gaseous envelopes and end type I migration 
\citep{pollack96, papaloizou05}. 

Co-orbital planets form as a natural consequence of gravitational scattering in
crowded systems, and occur in more than one
third of all models we have considered.
Simulations performed with sub-Earth mass protoplanets
showed that even very low-mass bodies could form and maintain
co-orbital systems.
We showed that an important factor in establishing these systems
was the existence of a relatively flat disc surface density
profile, which promotes convergence of planetary orbits
through differential type I migration. We also showed that
these co-orbital systems can be stable for over 3 billion years after
gas disc dispersal.
Co-orbital triples (i.e. three planets all in mutual 1:1 resonance)
are also potentially viable, though none were found
in the suite of simulations presented in this paper.
Previous $N$-body simulations, using a slightly modified form for
the eccentricity and damping formulae adopted in this paper, did
result in such systems occasionally forming \citep{cresswell06b}.
Although no planetary system is known to host such a
configuration, we note that 
such `tadpole twins' have been observed
in orbit around Saturn, with Helene and Polydueces occupying the $L_4$ and
$L_5$ points of Dione \citep{murray05}. 
The observation of co-orbital extrasolar planet systems
would be a strong indicator that a period of sustained
dynamical relaxation had occured during the formation of that system.

There are a number of open questions regarding the co-orbital
planets that form in our simulations. The main issue is whether
they can remain stable if one or both planets undergo gas
accretion to become giants, since such a system would be more
amenable to detection among the extrasolar planet population. 
Another question regarding our overall results is whether
a combination of planet-planet interactions and stochastic
migration, induced by turbulent density fluctuations in the disc
\citep{nelson04, nelson05},
can help prevent large-scale migration of some
planets within the swarm. In particular, we find that mean-motion
resonances are very effective in shielding the protoplanets from
close-encounters that lead to scattering, and stochastic torques
may increase the fragility of these resonances, including the co-orbitals.
Another point of interest is the effect of other halting mechanisms
for type I migration, such as a disc edge caused by a 
magnetospheric cavity \citep{terquem07} or a `planet trap'
produced by a region of positive density gradient in the disc
\citep{masset06, morbidelli08}, on a resonant group that has already formed and 
migrated some distance. \citet{terquem07} suggest that near-commeasurabilities
between protoplanets will survive beyond a disc edge, while experiments by 
\citet{pierens08} similarly indicate that a disc edge
formed by the action of a close stellar binary companion can prevent the
further migration of a moderate ($\sim5$) swarm of protoplanets, possibly 
subject to further reordering of the population.
Our results suggest that the survival of co-orbital systems in such 
circumstances is possible, but depends on the rate at which migration is 
halted and any further scattering among the system.
We will address these and other issues
relating to planetary swarms and co-orbitals in a future paper.

\acknowledgements
{The simulations reported here were performed using the UK Astrophysical
Fluids Facility (UKAFF) and the QMUL HPC facility purchased under the SRIF
initiative. We thank Doug McNeil for several interesting discussions on the
subject of co-orbital formation, and the anonymous referee for their
constructive comments}.

\bibliographystyle{aa}

\bibliography{3dbibliography}

\end{document}